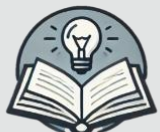

THE PATENTIST
LIVING LITERATURE REVIEW

# What Proportion of Knowledge is Patented?

Gaétan de Rassenfosse

Holder of the Chair of Science, Technology, and Innovation Policy
École polytechnique fédérale de Lausanne, Switzerland.

This version: January 2025

**Purpose**
This article is part of a Living Literature Review exploring topics related to intellectual property, focusing on insights from the economic literature. Our aim is to provide a clear and non-technical introduction to patent rights, making them accessible to graduate students, legal scholars and practitioners, policymakers, and anyone curious about the subject.

**Funding**
This project is made possible through a Living Literature Review grant generously provided by Open Philanthropy. Open Philanthropy does not exert editorial control over this work, and the views expressed here do not necessarily reflect those of Open Philanthropy.

# What proportion of knowledge is patented?

Gaétan de Rassenfosse
École polytechnique fédérale de Lausanne, Switzerland.

Knowledge is the lifeblood of innovation, fueling economic growth and improving living conditions. However, we don't treat all knowledge the same way. Some is patented, while other knowledge remains freely available or kept secret. Understanding the share of knowledge that is patented provides important insights into the functioning of the knowledge economy and the contexts in which the patent system effectively incentivizes innovation. If certain types of knowledge persistently fall outside the realm of patenting, one can hardly argue that patents should be credited with their creation. Exploring these issues deepens our understanding of how intellectual property shapes the creation and dissemination of knowledge.

**What is knowledge?**

Before attempting to quantify the proportion of knowledge that is patented, we need a clear working definition of knowledge. Knowledge is a complex concept interpreted differently across fields like philosophy or sociology. Information science gives us a useful definition from the viewpoint of economics. It introduces three components: data, information, and knowledge. Data refers to raw, unorganized facts. Information is processed and organized data that carries meaning. Knowledge, in turn, is the application of information in a specific context to enable decision-making.

Knowledge encompasses a vast spectrum of human understanding and experience, from practical skills to scientific theorems to cultural traditions. For example, Indigenous communities have observed plant growth cycles, animal behavior, and weather patterns over generations ('data'). They have come to recognize which plants have medicinal properties, understand animal migration routes, and predict seasonal changes ('information'). They can then use this information to cultivate crops, hunt effectively, and prepare for seasonal challenges ('knowledge'). Knowledge is embedded into all our actions, from riding a bike to cooking to applying a rule of thumb.

**What kind of knowledge does the patent system aim to encourage?**

The patent system provides an *economic incentive* to create new *technical* knowledge. The 'economic incentive' implies that the patent system channels inventive activities toward knowledge with significant economic value. However, it overlooks certain inventions that are highly useful but not economically valuable. To understand this, we need to think about the difference between use value and exchange value. Use value refers to the utility or practical usefulness of an object, whereas exchange value refers to what an object can be traded for in the market. Although goods with high use value may sometimes hold high monetary value, this is not always true. For instance, water is indispensable and possesses high use value, yet diamonds—despite being far less useful—command a much higher exchange value due to their rarity and demand. This phenomenon is known as Adam Smith's 'paradox of value.'

The term 'technical knowledge' must be understood in the context of the 'patentable subject matter,' which denotes inventions considered eligible for patent protection under the law. In the United States, [four categories of invention](#) are appropriate subject matter, including processes, machines, manufactures, and compositions of matter. The first category defines 'actions'—inventions that consist of a series of steps or acts to be performed—while the latter three categories define 'things' or 'products.' The law also explicitly excludes abstract ideas, laws of nature (like physical laws or biological processes), and naturally occurring substances, organisms, or phenomena from patentable subject matters.

In sum, the patent system incentivizes the creation of new, economically valuable technical knowledge. This knowledge is undoubtedly key to advancing social and economic progress. However, an equally important aspect of progress relies on other types of knowledge.

- First, 'old' knowledge—knowledge that is not new to the world and, therefore, not patentable—plays a vital role in economic growth. Think about the wheel—it was invented a long time ago, but we still use it in cars, bikes, and airplanes. Economists have long recognized that growth depends not only on creating new inventions but also on disseminating and applying existing ones across firms and societies.
- Second, not all useful knowledge is economically valuable. For instance, drugs targeting rare diseases, while highly beneficial, lack commercial appeal due to their small market size. Consequently, the patent system alone provides insufficient incentives for firms to develop such drugs.
- Third, useful knowledge is not always 'technical.' Think of management practices or organizational strategies that enhance efficiency and productivity. Moreover, scientific knowledge and discoveries, which fall outside the patent system, are also of immense importance.

**What proportion of patentable knowledge is patented?**

Now that we have more narrowly defined the type of knowledge that can be subject to patent protection, it becomes evident that the overwhelming majority of knowledge lies outside the scope of the patent system. While quantifying the exact proportion of knowledge that is patentable is a vain task, economists have focused on the more modest endeavor of estimating the proportion of patentable inventions that are actually patented. At first glance, one might assume that all patentable inventions would be protected by patents—after all, why would inventors forgo patent protection if they are entitled to it? However, as research on the 'propensity to patent' has shown, this intuition is misleading.

Estimating the exact percentage of inventions that are patented is challenging due to a lack of comprehensive data on all inventions. However, academic studies tend to show that patents are not the primary means of protecting inventions in most industries. Moser (2005) used data from nineteenth-century world fairs to estimate the proportion of inventions patented. She estimated that one in nine innovations presented at the 1851 [Crystal Palace Exhibition](#) in London were patented, and one in eight at the 1876 [Centennial Exposition](#) in Philadelphia. Patent law has strengthened significantly since then, increasing patenting incentives, and 'knowledge' has become a greater source of competitive advantage for firms, potentially increasing the urge to use patents.

Modern evidence about the 'propensity to patent' comes primarily from survey data. Mansfield (1986) surveyed 100 U.S. manufacturing firms in the early 1980s, asking them to estimate the percentage of their patentable inventions that were patented. He reported numbers ranging from 50 percent for the primary metal industry to 86 percent for the machinery industry. In a survey of 1478 labs in the U.S. manufacturing sector in the early 1990s, Cohen et al. (2001) obtained significantly lower numbers. Their respondents applied for patents on 49 percent of their product innovations and 31 percent of their process innovations. Arundel and Kabla (1998) provide a European perspective. They surveyed 604 of Europe's largest industrial firms in the early 1990s and reported that the average propensity to patent was about 36 percent for product innovations (but close to 80% in pharmaceuticals) and 25 percent for process innovations (and close to 50% in precision instruments).

Fontana et al. (2013) adopted a different approach to investigate patent propensity. They analyzed a sample of 2,800 inventions that received the 'R&D 100' award, a competition organized by the journal *Research and Development* (R&D) to recognize the 100 most technologically significant new products available for sale or licensing in the year prior to evaluation. Their dataset, spanning the period from 1977 to 2004, reveals that a mere 10 percent of these inventions were patented.

These results have considerably enriched our understanding, but they are based on relatively small samples. Mezzanotti and Simcoe (2023) address this limitation by providing estimates representative of the broader population of firms in the United States. Drawing on data from the U.S. Census Bureau's Business R&D and Innovation Survey, collected between 2008 and 2015, they estimate that about 80 percent of firms performing R&D activities—presumably leading to patentable inventions—do not apply for patents. In other words, patenting is not the norm but, rather, a privilege of a select few firms. However, the study also reveals that patenting firms account for the vast majority (above 90 percent) of reported R&D. Importantly, this finding does not imply that 90 percent of inventions are patented. Instead, it highlights that most inventions come from firms engaged in patenting, even if not every invention they create results in a patent application.

**Why are not all patentable inventions patented?**

The evidence presented above indicates that not all patentable inventions are, in fact, patented. While firms in some industries may seek patent protection for most or even all of their inventions, others patent selectively or choose not to patent at all. A body of literature in economics and management has explored the factors influencing firms' propensity to patent from both theoretical and empirical perspectives.

A key insight from this literature is the significant heterogeneity in patenting behavior across invention types, firm characteristics, industries, and market structures. While a comprehensive discussion of these factors is beyond the scope of this essay, it is worth noting that, in some cases, inventions eligible for patent protection are better kept secret or disclosed to the public without seeking IP protection.

To illustrate why firms may prefer secrecy over patenting, consider the 'enablement requirement' of patent law, which mandates that a patent application must disclose sufficient information to enable a person having ordinary skill in the art (PHOSITA) to replicate the invention. While patent protection would in practice prevent anyone from using the replicated inventions (at least in countries where patent protection is valid), the public disclosure of the invention's technical details can discourage firms from patenting. Instead, they may favor secrecy, particularly for inventions that are difficult or costly to imitate. Conversely, patenting becomes more attractive for inventions that are easy to reverse-engineer, such as mechanical designs or chemical formulas. This distinction helps explain why surveys consistently find that product inventions are more likely to be patented than process inventions. While products are often exposed to public scrutiny, manufacturing processes conducted behind closed doors can often remain effective trade secrets.

**Good news for the free use and sharing of knowledge?**

This review has highlighted that the vast majority of knowledge is not patentable, and even when it is, only a fraction becomes patented. Consequently, most knowledge remains free to use and share. However, a critical distinction must be made between using knowledge and sharing it. While patents restrict the use of patented knowledge, they promote its sharing by requiring detailed public disclosure. Patent databases have thus become invaluable repositories of technical information, freely accessible to all—a stark contrast to trade secrets, which, while freely usable if independently discovered, are rarely shared (except under non-disclosure agreements).

That said, even if only a small fraction of the world's knowledge is patented, the tensions arising from the exclusive rights of patents cannot be overlooked, particularly when they pertain to inventions of immense social importance. Pharmaceuticals exemplify these tensions, where patents often provide the financial incentive for innovation but can also delay access to life-saving treatments. Although patent protection is time-limited—typically 20 years from filing—this period can feel disproportionately long in critical situations, such as for life-saving treatment or during a global pandemic. The balance between incentivizing innovation and ensuring equitable access remains a pressing challenge for the IP system.